\documentclass[11pt,reqno]{amsart}
\usepackage{setspace}
\usepackage{epsfig}
\input epsf.tex
\usepackage{graphicx,epsfig}
\numberwithin{equation}{section}
\usepackage{amsmath}
\usepackage{amscd,amsthm,amsfonts,amsopn,amssymb,verbatim}
\usepackage{mathrsfs}
\numberwithin{equation}{section}
\theoremstyle{remark}

\def\be{\begin{equation}}
\def\ee{\end{equation}}
\def\vp{\varphi}
\def\ve{\varepsilon}

\allowdisplaybreaks
\parskip=\medskipamount
\begin{document}

\title
[A note on Lyapunov exponents]
{A note on Lyapunov exponents of deterministic strongly mixing potentials}
\author
{J.~Bourgain}
\address
{Institute for Advanced Study\\
1 Einstein Drive\\
Princeton, NJ 08540}
\author
{E.~Bourgain-Chang}
\address
{University of California, Berkeley}
\email
{ebc@berkeley.edu}
\begin{abstract}
In this Note, we consider 1D lattice Schr\"odinger operators with deterministic strongly mixing potentials as 
studied in \cite{C-S}, \cite{B-S} with very small coupling.
We describe a scheme to establish positivity of the Lyapunov exponent from a statement at some fixed scale.
The required input may then be derived from Furstenberg theory, if the underlying dynamics are sufficiently
mixing, or verified directly by numerical means.
\end{abstract} 
\maketitle

\section
{\bf Summary}

Consider the two following model  cases of Schrodinger operators on $\mathbb Z$
\be\label{1.1}
H_x =\Delta+V_x  \, \text { with } \, V_x(j) =\lambda f(K^jx), x\in\mathbb T, K\in\mathbb Z_+, K\geq 2 
\ee
and
\be\label{1.2}
H_x= \Delta+V_x \, \text { with } \, V_x(j)=\lambda f(A^jx), x\in\mathbb T^2, \, A\in SL_2(\mathbb Z) 
\text { hyperbolic}.
\ee

Here $f$ is a non-constant function that we assume at least $C^1$.

For small coupling $\lambda\not= 0$, positivity of the Lyapunov exponents was established in \cite {C-S} using the Figotin-Pastur
perturbative method.
This analysis was pursued in \cite {B-S} where we proved furthermore that the Integrated Density of States (IDS) is H\"older regular and $H_x$ satisfies Anderson
localization for almost every $x$.

At the other end, if  we assume for instance $f$ a non-constant trigonometric polynomial and $|\lambda|>\lambda_f$, M.~Herman
subharmonicity technique applies to prove positivity of the exponents (and hence Anderson localization using the arguments from \cite
{B-S}).

These seemed to be the only available strategies, which do not cover intermediate ranges of $\lambda$ (note also that Herman's method is
quite restrictive for the function $f$).

The purpose of this Note is to show how to prove the above properties (under mild assumptions on $f$) and arbitrary $\lambda\not= 0$,
provided the transformation is assumed `sufficiently mixing'.
More precisely, in \eqref{1.1} we assume $K$ large enough and in \eqref{1.2}, that $A$ is sufficiently expanding, depending on the oscillation of $f$
and $\Vert f\Vert_{C^1}$.
Eventually, positivity of the Lyapunov exponents is then derived from Furstenberg's theorem on random matrix products.

Precise statements appear in Theorems 1 and 2 below.

Roughly speaking, our approach relies on a finite scale statement.
This input is derived from large deviation principles in Furstenberg's theory, arguing that the cocycles at some fixed
finite scale behave like in the random case (for an appropriate assumption on the model).
Rather than invoking random matrix product theory, the required finite scale information may also be checked 
numerically, as we will illustrate in some simple cases.
We also believe that such `finitary approach' to problems of positivity of Lyapunov exponents could be 
useful in other situations.

\section{\bf Schrodinger cocycles}

Let $(\Omega, \mu, T)$ be a dynamical system, $\Omega =\mathbb T^k$ a torus, $\mu$ Lebesque.
This setting covers our examples.

Let $f$ be a $C^1$ function on $\Omega$.

Define
\be\label{2.1}
M_N(x)= \begin{pmatrix}E - f(T^{N-1} x) &-1\\ 1&0\end{pmatrix} \ \begin{pmatrix}E- f(T^{N-2} x)&-1\\ 1&0\end{pmatrix}\cdots
\begin{pmatrix} E-f(x)& -1\\ 1&0\end{pmatrix}.
\ee
Let $N_0$ be a fixed large scale and write
$$
M_{NN_0}(x) =M_{N_0} (T^{(N-1)N_0}x) M_{N_0} (T^{(N-2)N_0}x)\ldots M_{N_0}(x).
$$
Next, we perform a translation of $x$ by vectors $y\in \Omega_{N_0, N}\subset\Omega$ (where $\Omega_{N_0, N}$ may
be finite or infinite) and equipped with a normalized measure $\underset y {Av}$. Using again the cocycle property
\be\label{2.2}
\log\Vert M_{NN_0} (x+y)v\Vert = \log \Big\Vert M_{N_0} (T^{(N-1)N_0} (x+y)) \Big(\frac {M_{(N-1)N_0}(x)v}
{\Vert M_{(N-1)N_0}(x)v\Vert}\Big)\Big\Vert
\ee
\begin{align}\label{2.3}
&+\log \Vert M_{(N-1)N_0}(x)v\Vert \nonumber\\
&+\big[\log\Vert M_{NN_0}(x+y)v\Vert -\log \Vert M_{N_0}(T^{(N-1)N_0}(x+y)) M_{(N-1)N_0} (x) v\Vert\big].
\end{align}
Integrating, this gives
\begin{align}\label{2.4}
&\int\log\Vert M_{NN_0} (x)v\Vert dx\geq\nonumber\\
&\int\Bigg\{ \underset y{Av}  \, \log\Bigg\Vert M_{N_0} \big(T^{(N-1)N_0}(x+y)\big) \Big(\frac {M_{(N-1)N_0}(x) v}{\Vert M_{(N-1)N_0}(x)v\Vert}\Big)\Bigg\Vert\Bigg\} dx\\
&+\int\log \Vert M_{(N-1)N_0} (x)v\Vert dx\nonumber\\
&-(2.5)\nonumber
\end{align}
with (2.5) an upperbound on $|\eqref{2.3}|$ for $x\in \Omega$ and $y\in \Omega_{N_0, N}$.
\stepcounter{equation}

Obviously \eqref{2.4} is at least
\be\label{2.6}
\min_N\ \min_{x\in\Omega, |w|=1} \underset y {Av}\{\log \Vert M_{N_0} \big(T^{(N-1)N_0}(x+y)\big) w\Vert\}.
\ee

Our goal is then to ensure positivity of \eqref{2.6}, more precisely,
$$
\eqref{2.6}> (2.5).
$$
This process may then be iterated, leading to positivity of the Lyapunov exponent $L(E) =\lim_{N\to\infty}
\frac 1N\int_\Omega \log\Vert M_N(E, x)\Vert dx$.

The scale $N_0$ is chosen sufficiently large, depending on the dynamics and $f$.
The basic idea is to obtain an image measure on $\Omega^{N_0}$ under the map
\be\label{2.7}
y\mapsto \big(T^{(N-1)N_0}(x+ y), T^{(N-1)N_0+1}(x+y), \ldots, T^{NN_0-1}(x+y)\big)
\ee
which is close to the uniform measure on $\Omega^{N_0}$ (in the weak sense, depending on $f$).
This will require sufficiently strong mixing properties for the transformation $T$.
Eventually (assuming $f$ non-constant of course), our goal is to exploit the theory of random matrix
products, cf. \cite{B-L}.
This theory indeed asserts that by taking $N_0$ large enough,
\be\label{2.8}
\min_{|w|=1} \int_{\Omega^{N_0}} \log \Big\Vert \prod_{j=0}^{N_0-1} \begin{pmatrix} E-f(x_j)&-1\\ 1&0\end{pmatrix}
w\Big\Vert dx_0\cdots dx_{N_0-1} > cN_0
\ee
for $c\approx L(E)>0$ and \eqref{2.8} moreover valid for all $E$ in a given bounded energy interval
(note that for large $E$, clearly $L(E)\sim\log |E|)$.

It remains then to bound (2.5), which is an issue of cocycle perturbation and will typically depend on
expansion properties of $T$.
Some basic estimates will be presented in the next section.

There is the following variant of the above construction.
Taking \hfill\break $v\in \mathbb R^2, |v|=1$, one may write (denoting $M_N^*$ the adjoint of $M_N$)
\begin{align}\label{2.9}
\log \Vert M_{NN_0} (x+y) &\geq \log \Vert M^*_{NN_0} (x+y)v\Vert \geq \nonumber\\
& \ \log\big\Vert M_{N_0}^*(x+y) \frac {M^*_{(N-1)N_0} (T^{N_0}x)v}{\Vert M^*_{(N-1)N_0}(T^{N_0} x)v\Vert}\big\Vert\\
& +\log\Vert M^*_{(N-1)N_0} (T^{N_0}x)v\Vert\label{2.10}\\
-\max _{|v|=1, y\in\Omega_{N_0, N}} &\big|\log\Vert M^*_{NN_0} (x+y)v\Vert -\log \Vert M^*_{N_0} (x+y)
M^*_{(N-1)N_0} (T^{N_0}x)v\Vert \big|.\label{2.11}
\end{align}
Let then $v=v_x$ such that $\Vert M^*_{(N-1)N_0} (T^{N_0}x) v_x\Vert =\Vert M_{(N-1)N_0}(T^{N_0}x)\Vert$.
Hence
$$
\eqref{2.9} +\eqref{2.10} =\log \Vert M_{N_0}^* (x+y) w_x\Vert+\log 
\Vert M_{(N-1)N_0} (T^{N_0}x)\Vert
$$
for some unit vector $w_x$.
Integration in $x\in\Omega, y\in\Omega_{N_0, N}$ gives
\begin{align}\label{2.12}
\int\log\Vert M_{NN_0}(x)\Vert dx&> \min_{x\in\Omega,  |w|=1} \big\{ 
\underset y{Av}\log\Vert M^*_{(N_0)} (x+y)w\Vert\big\}\\
&+\int\log \Vert M_{(N-1)N_0} (x)\Vert dx\nonumber\\
&-\eqref{2.11}\nonumber
\end{align}
and we may iterate, provided \eqref{2.12}--\eqref{2.11} $>0$.

We will implement the above scheme in the following examples
\be\label{2.13}
\Omega=\mathbb T, Tx= Kx \text { with } \ K\geq 2 \text { an integer}
\ee
\be\label{2.14}
\Omega=\mathbb T^2, Tx=Ax \text { with $A\in SL_2(\mathbb Z)$ hyperbolic}.
\ee

\medskip

\section
{\bf Perturbation of cocycles}

Let $A_1, \ldots, A_N\in SL_2(\mathbb R); \Vert A_j\Vert, \Vert A_j^{-1}\Vert < C_1$

Let $B_1, \ldots, B_N\in SL_2(\mathbb R), \Vert A_j- B_j\Vert< \ve_j< 1$

Let $v\in\mathbb R^2, \Vert v\Vert =1$

The purpose of what follows is to establish inequality (3.11) below.

Estimate
$$
\begin{aligned}
&\Vert A_1 \ldots A_N v- B_1\ldots B_Nv\Vert\leq \Vert A_1-B_1\Vert \, \Vert A_2\ldots A_N v\Vert+ \Vert B_1A_2\ldots
A_Nv-B_1\ldots B_N v\Vert\\
&=(3.1)+(3.2)
\end{aligned}
$$
Then
\begin{enumerate}
\item [(3.1)] $\leq \ve_1\Vert A_2 \cdots A_Nv\Vert\leq \ve\Vert A_1^{-1} \Vert\, \Vert A_1 \cdots A_Nv\Vert < \ve_1 C_1\Vert A_1\ldots A_Nv\Vert$
\item [(3.2)] $\leq \Vert B_1\Vert \, \Vert A_2-B_2\Vert \, \Vert A_3\ldots A_N v\Vert +\Vert B_1B_2A_3\ldots A_Nv -
B_1 \ldots B_Nv\Vert = (3.3)+(3.4)$
\medskip
\item [(3.3)] $\leq (C_1+\ve_1)\ve_2 \Vert A_3 \ldots A_Nv\Vert \leq (C_1+1)\ve_2 \Vert A_2^{-1}\Vert \, \Vert A_1^{-1}\Vert\,
\Vert A_1 A_2\ldots A_Nv\Vert \break
< \ve_2 (1+C_1)C_1^2 \Vert A_1\ldots A_N v\Vert$
\medskip
\item [(3.4)] $\leq \Vert B_1\Vert\, \Vert B_2\Vert\, \Vert A_3-B_3\Vert \, \Vert A_4\cdots A_Nv\Vert + \Vert B_1B_2B_3 A_4\ldots A_Nv -B_1
\ldots B_Nv\Vert \break
= (3.5)+(3.6)$
\medskip
\item [(3.5)] $< (1+C_1)^2 \ve_3 \Vert A_3^{-1}\Vert \, \Vert A_2^{-1}\Vert\, \Vert A_1^{-1}\Vert \, \Vert A_1A_2\ldots A_N v\Vert
<\ve_3 (1+C_1)^2 C_1^3 \Vert A_1\ldots A_N v\Vert$
\end{enumerate}

etc.

Hence
$$
\Vert A_1\ldots A_Nv-B_1\ldots B_Nv\Vert <\Big[\sum_{j\geq 1} \ve_j(1+C_1)^{2j-1}\Big] \Vert A_1\ldots A_N v\Vert
\eqno{(3.7)}
$$
and also
$$
\Bigg\Vert\frac {A_1\ldots A_N v}{\Vert A_1\ldots A_N v\Vert} - 
\frac {B_1\ldots B_N v}{\Vert B_1\ldots B_N v\Vert}\Bigg\Vert
< 2 \sum_{j\geq 1} \ve_j(1+C_1)^{2j-1}.\eqno{(3.8)}
$$
Write a telescopic sum
$$
\begin{aligned}
\log\Vert A_1\ldots A_N v\Vert&=\log\Bigg\Vert A_1\Bigg(\frac {A_2\ldots A_Nv}{\Vert A_2\ldots A_N v\Vert}\Bigg)\Bigg\Vert\\[6pt]
&+\log \Bigg\Vert A_2\Bigg(\frac {A_3\ldots A_Nv}{\Vert A_3 \ldots A_N v\Vert}\Bigg)\Bigg\Vert\\[6pt]
&+\cdots\\[6pt]
&+\log\Bigg\Vert A_j\Bigg(\frac {A_{j+1} \ldots A_N v}{\Vert A_{j+1} \ldots A_Nv\Vert}\Bigg)\Bigg\Vert+\cdots
\end{aligned}
$$
By (3.8)
$$
\Bigg\Vert A_j\Bigg(\frac {A_{j+1}\ldots A_N v} {\Vert A_{j+1}\ldots A_N v\Vert}\Bigg) - 
B_j\Bigg(\frac {B_{j+1}\ldots B_n v} {\Vert B_{j+1} \ldots B_N v\Vert}\Bigg)\Bigg\Vert <\ve_j+2C_1\sum_{j'\geq j+1} \ve_{j'} (1+C_1)^{2{j'}-1}
$$
and therefore, since $\Vert B_j w\Vert\geq \Vert B_j^{-1} \Vert^{-1}>\frac 1{C_1}$ for $|w|=1$
$$
\left|\frac {\Vert A_j(\frac {A_{j+1}\ldots A_N v} {\Vert A_{j+1} \cdots A_N v\Vert})\Vert}
{\Vert B_j(\frac {B_{j+1}\ldots B_N v}{\Vert B_{j+1} \ldots B_N v\Vert})
\Vert} -1\right| < C_1\ve_j+2C_1^2 \, \sum_{j'\geq j+1} \ve_{j'} (1+C_1)^{2j'-1}.
$$
Hence, using the inequality $|\log (1+x)|< 2|x|$ for $|x|<\frac 12$, it follows that
$$
\begin{aligned}
&\left|\log\Big\Vert A_j\Big(\frac {A_{j+1}\ldots A_N v}{\Vert A_{j+1}\ldots A_Nv\Vert}\Big)\Big\Vert -\log \, 
\Big\Vert B_j\Big(\frac{B_{j+1}\ldots B_Nv}{\Vert B_{j+1} \ldots B_Nv\Vert}\Big)\Big\Vert\right|<\\[8pt]
&2C_1\ve_j +4C_1^2 \sum_{j'\geq j+1} \ve_{j'}(1+C_1)^{2j'-1}
\end{aligned}
\eqno{(3.9)}
$$
provided
$$
C_1^2 \sum_{j\geq 1} \ve_j(1+C_1)^{2j-1}<\frac 15.\eqno{(3.10)} 
$$
Assuming (3.10), it follows that
$$
\big|\log\Vert A_1\ldots A_N v\Vert -\log\Vert B_1\ldots B_Nv\Vert\big|
< 6C_1^2 \sum_{j\geq 1} j\ve_j(1+C_1)^{2j-1}.\eqno{(3.11)}
$$
\medskip

\section
{\bf Map $x\mapsto Kx $ on $\mathbb T$}

Let $\Omega =\mathbb T, K\in\mathbb Z_+, K\geq 2$ and $f\in C^1(\mathbb T)$.

We apply the second procedure discussed in Section 2.

Fix $N_0$ and define $\Omega_{N_0, N} =\Omega_{N_0} =\Big\{\frac j{K^{N_0}}; j=0, 1, \ldots, K^{N_0}-1\Big\}$,
noting that
$$
T^{N_0} (x+y) \equiv T^{N_0} (x) (\text {mod 1) for all $x\in\mathbb T, y\in\Omega_{N_0}$}.
$$
Hence \eqref {2.11} $=0$ and \eqref{2.12} equals
\begin{align} \label{4.1}
&\min_{x, |w|=1} \Big\{ K^{-N_0} \sum_{\alpha =0}^{K^{N_0}-1} \log 
\Big\Vert \begin{pmatrix} E-f(x+K^{-N_0}\alpha) &1\\ -1 & 0\end{pmatrix}\nonumber\\
&\begin{pmatrix} E-f(Kx+K^{-N_0+1}\alpha)&1\\ -1&0\end{pmatrix} \cdots
\begin{pmatrix} E-f(K^{N_0-1}x + \frac \alpha K)  &1\\ 1&0\end{pmatrix} \begin{pmatrix} w_1\\ w_2\end{pmatrix}\Big\Vert\Big\}
\end{align}

Remains to analyze the map $\vp =\vp_x:\mathbb Z/ K^{N_0}\mathbb Z \to \mathbb T^{N_0}$ defined by
\be\label{4.2}
\vp(\alpha) =\Big(x+\frac\alpha{K^{N_0}}, Kx+\frac \alpha{K^{N_0-1}}, 
\ldots, K^{N_0-1} x+\frac \alpha K\Big).
\ee
If we fix $N_0$ and take $K$ large enough, the image measure becomes weakly equidistributed (uniformly in $x$).
One has indeed that for $\xi\in\mathbb Z^{N_0}$, \hfill\break $0<|\xi|< K$,
$$
K^{-N_0}\Big|\sum_{\alpha =0}^{K^{N_0}-1} e^{2\pi i\xi.\vp(\alpha)}\Big|= K^{-N_0}
\Big|\sum_{\alpha=0}^{K^{N_0}-1}  \ e^{2\pi i\alpha(\frac {\xi_0}{K^{N_0}} +\frac {\xi_1}{K^{N_0-1}}
+\cdots+ \frac {\xi_{N_0-1}} K)} \Big|=0
$$
implying the required equidistribution statement.

In view of the discussion in Section 2, we proved the following

{\bf Theorem 1.} {\it Given $\kappa >0$ and $0<C<\infty$, there is $K_0$ such that if $K\in\mathbb Z_+$, $K>K_0$, and $f\in C^1(\mathbb T)$ is a real function satisfying
\be\label{4.3}
\Vert f\Vert_{C^1} <C \, \text { and } \, osc(f) >\kappa
\ee
then the Schrodinger operator on $\mathbb Z_+$
\be\label{4.4}
H_x =\Delta+V_x \, \text { with } \, V_x(j)=f(K^jx)
\ee
has positive Lyapunov exponents for all energies.}

Once the positivity of the Lyapunov exponents established, we may proceed further, following \cite {B-S}, and prove a large deviation inequality for the
pointwise Lyapunov exponents
$$
L_N(E; x) =\frac 1N \log\Vert M_N(E; x)\Vert.
$$
This enables then to establish H\"older regularity of $L(E)$ and the IDS $N(E)$ (by Thouless formula) 
and also Anderson localization (see \cite {B-S} for details).

Recall that the large deviation estimate required in this analysis is of the form
\be\label{4.5}
\text{mes\,}[x\in\mathbb T; |L_N(E, x) -L_N(E)|> o(1) L_N(E)]< e^{-cN}
\ee
where $L_N(E)=\int_{\mathbb T} L_N(E, x) dx$.

We indicate a proof of \eqref{4.5} by elaborating upon the above considerations.

Denote $\prod$ the product space $\prod_{\mathbb Z_+}\{0, 1\ldots, K^{N_0}-1\}$ equipped with normalized product
measure.
Given $\alpha=(\alpha_1, \alpha_2, \ldots)\in\prod$, perform a shift $x\mapsto x+\sum_{j\geq 1} \, \frac {\alpha_j}
{K^{jN_0}}$.
Clearly $M^*_{NN_0} (x+\sum_{j\geq 1} \, \frac {\alpha_j}{K^{j N_0}})=$
$$
 M^*_{N_0} \Big (x+\sum_{j\geq 1} \, \frac{\alpha_j}{K^{jN_0}}\Big)
 M^*_{N_0} \Big( K^{N_0} x+\sum_{j\geq 2} \frac {\alpha_j}{K^{(j-1)N_0}}\Big)\ldots
M^*_{N_0} \Big(K^{(N-1)N_0} x+\sum_{j\geq N} \frac {\alpha_j}{K^{(j-N+1)N_0}}\Big)
$$
and, fixing $x\in \mathbb T$, $w\in\mathbb R^2, |w|=1$.
\begin{align}\label {4.6}
&\log\Big\Vert M_{NN_0}^* \Big(x+\sum_{j\geq 1} \, \frac {\alpha_j}{K^{jN_0}}\Big) w\Big\Vert =\nonumber\\
&\log\Big\Vert M_{N_0}^* \Big(x+\sum_{j\geq 1} \, \frac {\alpha_j}{K^{jN_0}}\Big) w_1 (\alpha_j; j\geq 2)\Vert +\nonumber\\
&\log\Big\Vert M_{N_0}^*\Big(x+\sum_{J\geq 2} \, \frac {\alpha_j}{K^{(j-1)N_0}}\Big) w_2 (\alpha_j; j\geq 3)\Big\Vert
+\cdots+\nonumber\\
&\log \Big\Vert M^*_{N_0} \Big(x+\sum_{j\geq N} \, \frac {\alpha_j}{K^{(j-N+1)N_0}}\Big) 
w_N(\alpha_j; j\geq N+1)\Big\Vert
\end{align}
where $w_1, w_2, \ldots $ are unit vectors in $\mathbb R^2$.

Rewrite the sum $\vp_1+\vp_2 +\cdots+ \vp_N$ in \eqref{4.6} as
$$
\begin{aligned}
& d_1+d_2+\ldots +d_N\\
&+\\
&\mathbb E[\vp_1|\alpha_1] +\cdots+ \mathbb E[\vp_N|\alpha_N]
\end{aligned}
$$
where $\{d_j=\vp_j -\mathbb E[\vp_j|\alpha_j]\}$ is a martingale difference sequence wrt the filtration
$\prod$ introduced above.  Also
$$
\begin{aligned}
&\max_{x\in\mathbb T,|w|=1} \ \underset{\mathbb Z/K^{N_0}\mathbb Z} {Av}\log \Big\Vert M^*_{N_0} \Big(
x+\frac \alpha{K^{N_0}}\Big)w\Big\Vert \geq\\
& \qquad \mathbb E[\vp_j| \alpha_j]\geq\\
&\min_{x\in \mathbb T, |w|=1} \underset{\mathbb Z/K^{N_0}\mathbb Z}{Av} \log \Big\Vert M^*_{N_0} \Big(
x+\frac \alpha{K^{N_0}}\Big)w\Big\Vert.
\end{aligned}
$$

If we fix $N_0$ and take $K$ large enough, then, uniformly in $x$ and $w$,
\begin{align}\label {4.7}
&\underset{\mathbb Z/K^{N_0}\mathbb Z} {Av} \log \Big\Vert M_{N_0}^* \Big(x+\frac \alpha{K^{N_0}}\Big)w\Big\Vert \approx\nonumber\\
&\int_{\mathbb T^{N_0}} \log\Big\Vert\begin{pmatrix} E-f(x_0)& 1\\ -1& 0\end{pmatrix}\ldots
\begin{pmatrix} E-f(x_{N_0-1})&1\\ -1&0\end{pmatrix} w\Big\Vert dx_0 \cdots dx_{N_0-1}
\end{align}
as pointed out above, while, provided $N_0$ is chosen large enough, random matrix product theory implies that
$$
\frac 1{N_0} \eqref{f} \asymp L(E)>0 \ (= \text { Lyapunov exponent of the random cocycle)}.
$$
Hence, we get
\be\label{4.8}
\Big|L_{NN_0} \Big(E; x+\sum_{j\geq 1} \, \frac {\alpha_j}{K^{jN_0}}\Big) 
-L(E)\Big| < o(1)+\frac 1{NN_0} \Big| \sum^N_{j=1} d_j\Big|.
\ee
Applying the large deviation estimate for martingale difference sequences in the variable $(\alpha_j)\in\prod$,
it follows that
\be\label{4.9}
\text{mes\,}\Big[\alpha; \Big|\sum _{j=1}^N d_j\Big|> \delta NN_0\Big] < e^{-\delta' NN_0}
\ee
for some $\delta'= \delta' (\delta)>0$.
Therefore also
\be\label{4.10}
\text{mes\,}[x\in\mathbb T; | L_{NN_0} (E;x)-L(E)|>\delta] < e^{-\delta'N_0N}
\ee 
proving \eqref{4.5}.

\noindent
{\bf Remark 1.}

Take $f=\lambda f_0$ with $f_0$ satisfying \eqref{4.3} and let $\lambda$ vary.
Theorem 1 then applies in any fixed range $0<\lambda_1\leq |\lambda|\leq \lambda_2$ (with a same $K$).
For large $|\lambda|$ one easily derives positivity of \eqref{4.1} already for $N_0=2$.
On the other hand, the small $\lambda$ case is captured by the Figolin-Pastur perturbative method
(see \cite{C-S}, \cite{B-S} ) provided $F$ is restricted to $[-2+\delta, -\delta] \cup [\delta, 2-\delta]$ for
some $\delta>0$.

\noindent
{\bf Remark 2.}

For given $f$, rather than deriving positivity of \eqref{4.1} for appropriately chosen large $N_0, K$ by
invoking Furstenberg's theorem, one may of course proceed by a direct numerical verification at some scale $N_0$
and  any given $K$.
Hence positivity of the Lyapunov exponent for the model $x\mapsto Kx$ with given $f$ may in principle be
established numerically.
We will illustrate this with some examples at the end of the paper.
\medskip

\section
{\bf Toral automorphisms}

Let $\Omega=\Bbb T^2$ and $A=\begin{pmatrix} a&b\\c&d\end{pmatrix} \in SL_2(\mathbb Z)$ acting on $\mathbb T^2$ (the approach works similarly in the higher dimensional case).

We assume $A$ strongly expanding with a large expanding eigenvalue \hfill\break $K=\lambda_+ \sim t=a+d$ and expanding eigenvector $v=v_+\approx \frac {(a,
c)}{V\overline{a^2+c^2}}$.

Assume also $\Vert A\Vert\sim K$.

Returning to the discussion in Section 2, we apply the first scheme taking for $\Omega_{N_0, N} =\{K^{-(N-1)N_0+\frac 12} tv; 0\leq t\leq 1\}$.
Thus \eqref{2.2} becomes
\be\label{5.1}
\log\Vert M_{N_0} (A^{(N-1)N_0} x+ tK^{\frac 12} v) (w_x)\Vert
\ee
with
\be\label{5.2}
w_x= \frac {M_{(N-1)N} (x) v} {\Vert M_{(N-1)N_0}(x) v\Vert}.
\ee
In the present situation, the error terms (2.5) do not vanish and will be evaluated using the estimates
from Section 3.

Clearly
\be\label{5.3}
\Vert M_{N_0}\Vert, \Vert M_{N_0}^{-1} \Vert\leq (\Vert f\Vert_\infty + E+2)^{N_0} =C_2^{N_0}= C_1
\ee
and
\be\label{5.4}
\Vert M_{N_0} (x+A^\ell y)-M_{N_0}(x) \Vert \leq 2C_1 \Vert f'\Vert_\infty K^{N_0+\ell -1} |y|<
2C_1 \Vert f'\Vert_\infty K^{-(N-2)N_0+\ell -\frac 12}.
\ee
For $j=1, \ldots, N$, define
$$
A_j=M_{N_0} \big(T^{(N-j)N_0} (x+y)\big) = M_{N_0} (A^{(N-j)N_0} x +A^{(N-j)N_0} y)
$$
and
$$
B_j=M_{N_0} (T^{(N-j)N_0} x) =M_{N_0} (A^{N-j)N_0}x).
$$
Applying \eqref{5.4} with $\ell =(N-j)N_0$ implies
$$
\Vert A_j-B_j\Vert < 2C_1\Vert f'\Vert_\infty \ K^{-(j-2) N_0-\frac 12}=\ve_j.
$$
Condition (3.10) becomes
$$
2C_1^3 \Vert f'\Vert_\infty \sum_{j\geq 2} K^{-(j-2)N_0-\frac 12} (1+C_1)^{2j-1} < \frac 15
$$
which by \eqref{5.3} will be satisfied if
\be\label{5.5}
K> 10^3 C_1^6 (1+C_1)^6 (1+\Vert f'\Vert_\infty)^2.
\ee
An application of (3.11) gives then
$$
|\eqref {2.3}|, |(2.5)|< 6 C_1^2 \sum_{j\geq 2} j \ve_j (1+C_1)^{2j-1} < 20 C_1^3 (1+C_1)^3\Vert f'\Vert_\infty K^{-\frac 12}
$$
which can be made arbitrarily small by choosing $K$ large.

Remains to consider the image measure under the map \eqref{2.7}
\begin{align}\label{5.6}
 \vp: &[0, 1] \to (\mathbb T^2)^{N_0}: t\mapsto (A^{(N-1)_{N_0}} x+K^{\frac 12} tv, A^{(N-1)N_0+1} x\nonumber\\
&+ K^{\frac 32 } tv, \ldots , A^{NN_0-1} x+ K^{N_0-\frac 12} tv)
\end{align}
or, equivalently
\be\label {5.7} 
 \psi : [0,1]\to (\mathbb T^2)^{N_0}: t\mapsto (K^{\frac 12} tv, K^{\frac 32} tv, \ldots, K^{N_0-\frac 12} tv).
\ee
Estimating the Fourier transform of the image measure $\eta$ of $\psi$ on $(\mathbb T^2)^{N_0}$, we get
for $(\xi_0, \ldots, \xi_{N_0-1})\in (\mathbb Z^2)^{N_0}$
\begin{align}\label{5.8}
|\hat\eta(\xi_0, \ldots, \xi_{N_{0-1}})| & = \Big|\int^1_0 
e\big(K^{\frac 12} ( v.\xi_0+Kv.\xi_1+\cdots + K^{N_0-1} v.\xi_{N_0-1})t\big)dt\Big|<\nonumber\\
& 4[1+K^{\frac 12} |v.\xi_0+Kv.\xi_1+\cdots+ K^{N_0-1} v.\xi_{N_0-1}|]^{-1}.
\end{align}
We restrict ourselves to frequencies $(\xi_0, \ldots, \xi_{N_0-1})\in(\mathbb Z^2)^{N_0}$ with $|\xi_j|<B=B_f$.

We assume $A$ sufficiently mixing, in the sense that the expanding vector $v$ satisfies a diophantine property
\be\label{5.9}
|\langle v, \xi\rangle|> \frac 1{B_1} \ \text{ for all } \ \xi\in \mathbb Z^2\backslash \{0\}, |\xi|< B.
\ee
Also assume $K=\lambda_+> 2BB_1$.
It follows then from \eqref{5.9} that \eqref{5.8} $<\frac {4B_1}{K^{\frac 12}}$ if $(\xi_0 , \ldots, \xi_{N_0-1}) \in
(\mathbb Z^2)^{N_0}\backslash \{0\}$, $|\xi_j|<B$.
We obtain 
\medskip

\noindent{\bf Theorem 2.}
{\it Consider the Schrodinger operator on $\mathbb Z$
\be\label{5.10}
H_x=\Delta+\lambda V_x\text { with } V_x(j)=f(A^jx)
\ee
with $f$ a non-constant function in $C^1(\mathbb T^2)$, $A\in SL_2(\mathbb Z)$.
Assuming $A$ sufficiently mixing (depending on $osc(f|\mathbb T^2$) and $\Vert f\Vert_{C^1}$),
we obtain positive Lyapunov exponents for all energies, H\"older regularity of the IDS and Anderson localization.
}
\medskip

\section
{\bf Further comments and numerical aspects}

As pointed out already, the required positivity of \eqref{2.6} at some scale $N_0$ may be verifiable numerically,
short of a theoretical reason, and in this way extensions of Theorems 1 and 2 could be obtained in  situations
where a former transference to a random setting is not possible.
Observe also that in our problem, one may replace $M_N(x)$ by $SM(x)S^{-1}$ for any chosen similarity $S\in SL_2(\mathbb R)$
such that $\Vert S\Vert<C$ independently from $N$.
Therefore, instead of establishing positivity of \eqref{2.6}, we can as well consider
\be\label{6.1}
\min_N \, \min_{x\in\Omega, |w|=1} \underset y {Av} \{\log\Vert SM_{N_0} \big(T^{(N-1)N_0} (x+y)\big) S^{-1} w\Vert\}
\ee 
for a {\it fixed} $S\in GL_2(\mathbb R)$.
In particular, using the Figolin-Pastur formalism, one may represent the Schr\"odinger matrices in polar coordinates 
(which is especially useful in the small $\lambda$ regime).

Fix $\delta >0$ and assume
\be\label{6.2}
|E|<2-\delta.
\ee
Denoting $v_n =f(T^n x)$, define $\kappa\in (0, \pi)$ and $V_n$ by
\be\label{6.3}
E= 2\cos \kappa
\ee
\be\label{6.4}
V_n = -\frac {v_n}{\sin \kappa}
\ee
and let
\be\label{6.5}
S=\begin{pmatrix} 1& -\cos \kappa\\ 0& \sin \kappa\end{pmatrix}.
\ee
Then
$$
M_N=\prod^0_{n=N-1} \begin{pmatrix} E-\lambda v_n&-1\\ 1&0\end{pmatrix}
$$
is converted to
\be\label{6.6}
M_N' =\prod^0_{n=N-1} \left[ \begin{pmatrix} \cos \kappa &- \sin\kappa\\ \sin \kappa& \cos \kappa\end{pmatrix}+
\lambda V_n\begin{pmatrix} \sin \kappa &\cos\kappa\\ 0&0\end{pmatrix}\right].
\ee

One should also expect that for small $\lambda$, the representation \eqref{6.6} is more suited for numerics as the
factors are perturbations of a rotation.

As an example related to \S4 of such numerics, we considered the map $x\mapsto 2x(K=2)$ and evaluated \eqref{4.1}
in the energy range $[-\sqrt 2, \sqrt 2]$ for different couplings $\lambda$ and using the transformation \eqref{6.6}.
It turns out that $N_0=6$ already suffices to obtain the positivity for $|\lambda|\geq \frac 25$.

In the displays below, we dropped the irrelevant factor $K^{-N_0}$ in \eqref{4.1}.

Further numerical work on the positivity of the Lyapunov exponent using our method will appear in the arXiv version of the
paper.

\begin{figure}[ht]
\centerline{
{\includegraphics[scale=0.75]{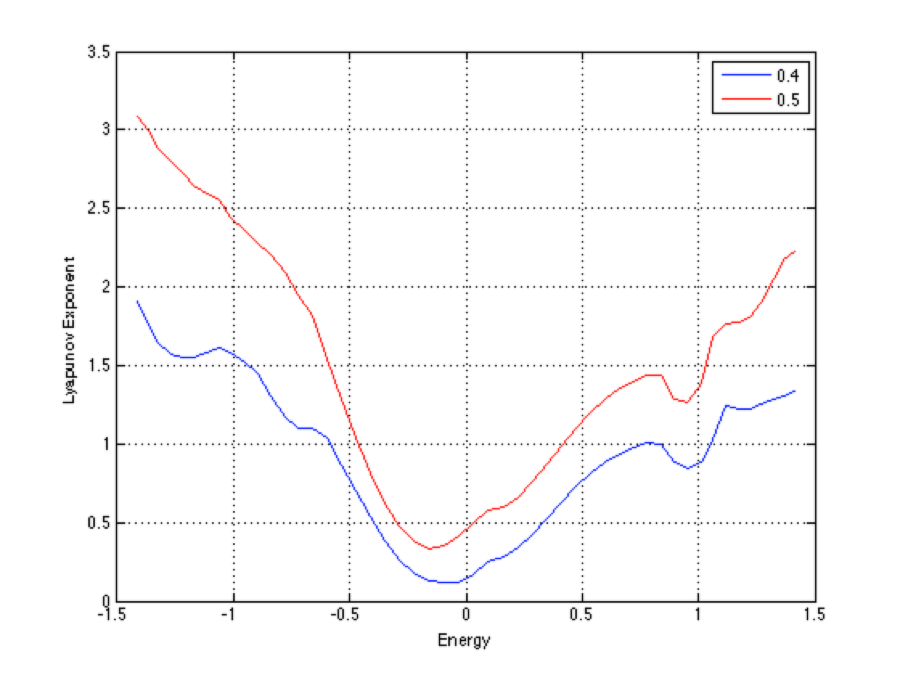}}}
\label{Fig. 1}
\caption{Positivity of Lyapunov Exponents for $\lambda=$ 0.4 and 0.5}
\end{figure}
\begin{figure}[ht]
\centerline{
{\includegraphics[scale=0.75]{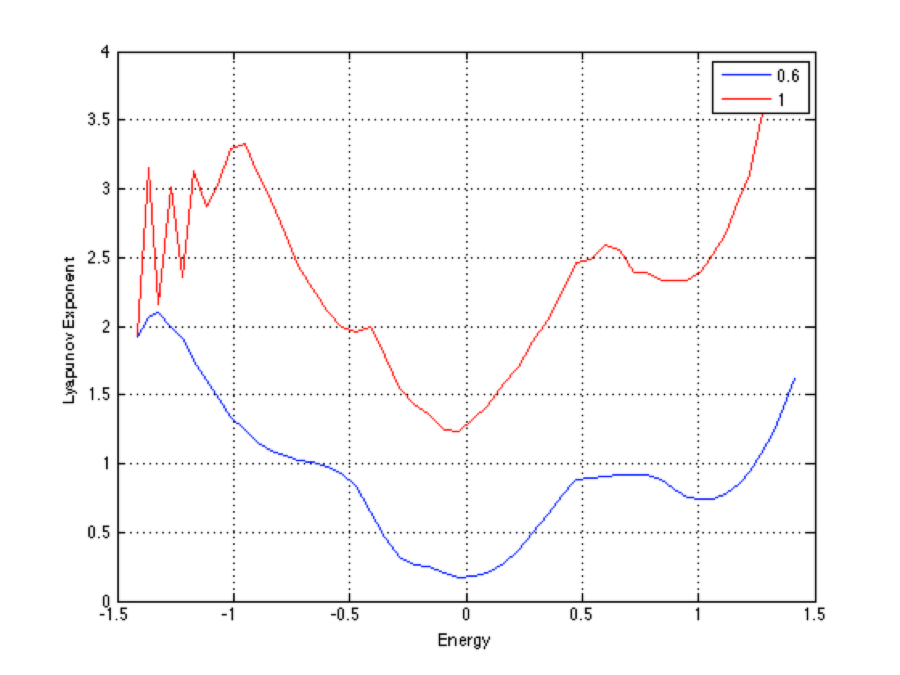}}}
\caption{Positivity of Lyapunov Exponents for $\lambda=$ 0.6 and 1}
\label{Fig. 2}
\end{figure}
\begin{figure}[ht]
\centerline{
{\includegraphics[scale=0.75]{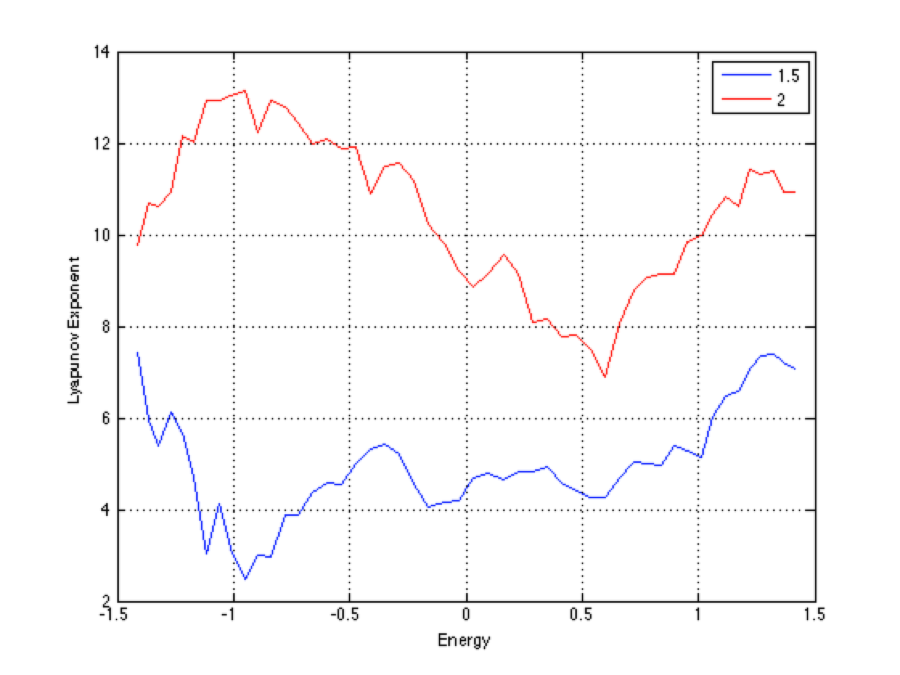}}}
\caption{Positivity of Lyapunov Exponents for $\lambda=$ 1.5 and 2}
\label{Fig. 3}
\end{figure}

\end{document}